\documentclass[onecolumn,showpacs,preprintnumbers,amsmath,amssymb]{article}

\usepackage{graphicx}
\usepackage{dcolumn}
\usepackage{bm}
\usepackage{latexsym,amsmath,amsthm,amssymb,amsfonts}
\usepackage{xcolor}

\usepackage{geometry}
\usepackage[affil-it]{authblk}

\title{Fluid and scalar field representations in the Brans-Dicke theory: cosmological scenarios}

\author[1]{Tiago H. B.  Alves}
\author[1]{Júlio C. Fabris}
\author[1,2]{Luiz Filipe Guimarães}

\affil[1]{Núcleo Cosmo-Ufes \& Departamento de Física, UFES, Vitória, ES, Brazil}
\affil[2]{{Departamento de F\'isica, Universidade Estadual de Londrina, Rod. Celso Garcia Cid, Km 380, 86057-970 Londrina, PR, Brazil}}

\begin{document}

\maketitle

\begin{abstract}

The Brans-Dicke scalar-tensor cosmological models are studied in both Einstein and Jordan frames, using hydrodynamical and self-interacting scalar field representations of the
energy-momentum tensor, leading to the same background solutions. The main features of the corresponding cosmological scenarios are determined. In many cases, we identify a transition from decelerated to accelerated regimes. The properties of the self-interacting scalar field, including the corresponding potential, are determined. Finally, some possible realistic configurations are analyzed.

\end{abstract}

\section{Introduction}

The $\Lambda$CDM cosmological model is very successful in reproducing the different observational data \cite{mukha}. However, this is done by introducing a dark sector containing two extra components in the cosmic matter/energy content: dark matter, a fluid with null (or almost null) effective pressure; and dark energy, which is generally connected with the quantum vacuum state.  Dark matter is related to the structure formation process and the dynamics of local structures, while dark energy is responsible for the present stage of expanding acceleration of the universe. There are many indirect evidences for the existence of these two dark components, but no direct detection, despite many efforts \cite{dd,de,bertone}. As a consequence, the nature of dark matter and dark energy remains unknown. This fact leads to many proposals to represent either dark matter and dark energy. Moreover, at least in what concerns dark energy, it has been evoked the possibility that the our standard gravitational theory, the General Relativity theory, must  be modified at large scales, while keeping its standard results at small scales. For a review, see for example \cite{mg} and references therein.

Since the nature of the components of the dark sector is unknown, there are many proposals to describe them either through fluids with a given equation of state or fields, specially self-interacting scalar fields. Many of these proposals are justified via fundamental frameworks (grand unified theory, string theory, supersymmetry, etc.),while many others are purely phenomenological. Among the different formulations of the dark sector we can also mention the attempt of incorporating both components in a single one. Chaplygin gas is one example \cite{ugo}, and it can be implemented either in a fluid representation or as a scalar field with a non-standard kinetic term that resembles the non-linear electrodynamics formulation. General discussions on the correspondence fluid/self-interacting scalar field can be found in Ref. \cite{madsen,vf0}.

The formal equivalence of  the description of the matter/energy sources, for a given cosmological model, through a fluid or a more fundamental field framework, is generally assured  for a high symmetric (isotropic and homogeneous) unperturbed universe. However, if we turn to the formation of the observed structures in the universe, the results depend crucially if a hydrodynamical or field representation is used, mainly at small scales. If adiabatic perturbations are used, for example, the sound velocity is given, in the context of a fluid, by the equation of state parameters. If the pressure is negative (as it is required by the dark energy component) the sound velocity becomes imaginary and the configuration is unstable specially due to the small scales modes \cite{jerome}. If the cosmic component is described by a field, the velocity of sound associated with the perturbations depends on the nature of the field. For a canonical self-interacting scalar field, for example, the sound speed is always equal to 1 in units of $c$, and the instabilities that may be present in the fluid description disappear. For a non-canonical scalar field, the scenario may be more convoluted. For example, in the case of a k-essence scalar field, with a non-canonical kinetic term for the scalar field, the hydrodynamical results are recovered. For other non-canonical scalar field structures the associated velocity of sound may be much more complicated. For a discussion on this issue, see Ref. \cite{neven}. 

In Ref. \cite{Tomi}, the fluid and self-interacting scalar field representations have been discussed in the context of cosmological models arising from the General Relativity (GR) theory. The goal of the present article is to analyze the fluid and self-interacting scalar field representation in scalar-tensor theory of gravity, in occurrence, the Brans-Dicke theory \cite{bd} - perhaps the simplest non-trivial class of scalar-tensor theories. Scalar-tensor theories present a direct (non-minimal) coupling between the scalar field and the curvature term. This is usually referred to as the Jordan frame (JF). This coupling can be reformulated as a minimal one, in a structure more similar to GR, through a conformal transformation, leading to the Einstein frame (EF) framework. This transformation simplifies the equations, but it leads to a non-trivial coupling of the scalar field in the matter sector. There are many discussions in the literature about the meaning of one or other formulation of a scalar theory, see for example Ref. \cite{vf}. In general, the conformal transformation leads to different frameworks, with different descriptions of the same physical system. For example, geodesic trajectories in the JF are mapped into non-geodesic trajectories in the EF. The description of the cosmic evolution, as well as that of the perturbed quantities on it, may appear quite different when passing from one frame to another.

Some general cosmological solutions of BD theory, for a flat universe and using a fluid description for the matter sector, determined in Ref. \cite{gurevich}, will be explored to construct the field representation. The analysis will be made both in the Jordan and Einstein frames. Due to the complexity of the background solutions, mainly the asymptotic solutions will be considered since they allow a direct mapping from one frame to another.
In both frames we aim to settle out the properties of the fluid and self-interacting scalar field representations in the matter sector. It will be shown, for example, that an accelerated expansion in one frame may be mapped into a decelerated expansion in another frame. In special, some classes of BD cosmological solutions admit an interpolation of decelerated and accelerated expansion, similarly to what happens in unified formulations of the dark sector, such as the Chaplygin gas proposal. Some of the scenarios representing the decelerated/accelerated transition may represent a realistic configuration in the sense that it can satisfies some other constraints like PPN\cite{will,ppn}. However, other configurations not satisfying these constraints may have sense in a more general scalar-tensor theory with dynamical couplings terms which may depend on the scales.

The nature of the attractive and repulsive behavior of the matter component will also be addressed. The analysis will be made more specific for three cases: zero pressure, cosmological constant and a phantom component. In the case a self-interacting scalar field is employed, the properties of the potential in each case, and in each frame, will be discussed. This analysis prepares the ground for a future complete perturbative study of cosmological scenarios, both in the JF and in the EF.

In section 2 the hydrodynamical and self-interacting formulations of a given solution in the JF are discussed. In section 3 the same analysis is performed in the EF. The results are analyzed in more details, both in the JF and in the EF, in section 4. In section 5 we present our main conclusions.

\section{Basic equations of the Brans-Dicke Theory}

The original Brans-Dicke theory is defined by the Lagrangian \cite{bd},
\begin{eqnarray}
\label{lm}
{\cal L} = \sqrt{-g}\biggr\{\Phi R - \omega\frac{\Phi_{;\rho}\Phi^{;\rho}}{\Phi}\biggl\} + \sqrt{-g}{\cal L}_m,
\end{eqnarray}
where $R$ is the Ricci scalar, $\Phi$ is a scalar field connected with the (inverse) of a the dynamical gravitational coupling, and ${\cal L}_m$ indicates the Lagrangian of the matter component, while $\omega$ is a coupling parameter. In general, RG is recovered in the limit $\omega \rightarrow \infty$ and $\Phi =$ constant, except for some special cases which includes radiation ($p = \rho/3$) and stiff matter fluids ($p = \rho$) \cite{cr,gb}. Units with $c = 1$ are used.

The equations of motion are,
\begin{eqnarray}
R_{\mu\nu} - \frac{1}{2}g_{\mu\nu}R &=& \frac{8\pi }{\Phi}T_{\mu\nu} + \frac{\omega}{\Phi^2}\biggr\{\Phi_{;\mu}\Phi_{;\nu} - \frac{1}{2}g_{\mu\nu}\Phi_{;\rho}\Phi^{;\rho}\biggl\} \nonumber \\
&+& \frac{1}{\Phi}\biggr\{\Phi_{;\mu;\nu} - g_{\mu\nu}\Box\Phi\biggl\},\\
\Box\Phi &=& \frac{8\pi}{3 + 2\omega}T\\
{T^{\mu\nu}}_{;\mu}  &=& 0.
\end{eqnarray}
The energy-momentum tensor is defined from the variational principle as,
\begin{eqnarray}
T_{\mu\nu} = - \frac{2}{\sqrt{-g}}\frac{\delta( \sqrt{-g}{\cal L}_m)}{\delta g^{\mu\nu}}.
\end{eqnarray}
When the energy-momentum tensor is described by the perfect fluid representation, it reads,
\begin{eqnarray}
T^{\mu\nu} = (\rho + p)u^\mu u^\nu - g^{\mu\nu}p,
\end{eqnarray}
where $\rho$ is the matter/energy density and $p$ is the corresponding pressure.

If the source is a self-interacting scalar field, the Lagrangian reads,
\begin{eqnarray}
\label{lo}
{\cal L} = \sqrt{-g}\biggr\{\Phi R - \omega\frac{\Phi_{;\rho}\Phi^{;\rho}}{\Phi} - \epsilon\phi_{;\rho}\phi^{;\rho} + 2V(\phi)\biggl\},
\end{eqnarray}
the corresponding energy-momentum tensor of a scalar field is given by,
\begin{eqnarray}
	8\pi T_{\mu\nu} = \epsilon\biggr(\phi_{;\mu}\phi_{;\nu} - \frac{1}{2}g_{\mu\nu}\phi_{;\rho}\phi^{;\rho}\biggl) + g_{\mu\nu}V(\phi).
\end{eqnarray}
In these expressions, $\epsilon = \pm$, in order to cover the scalar field satisfying the null energy condition (ordinary scalar field, plus sign) and violating the null energy condition (phantom scalar field, minus sign).

The field equations for a self interacting scalar field source are given by,
\begin{eqnarray}
R_{\mu\nu} - \frac{1}{2}g_{\mu\nu}R &=& \frac{1}{\Phi}\biggr\{\epsilon\biggr(\phi_{;\mu}\phi_{;\nu} - \frac{1}{2}g_{\mu\nu}\phi_{;\rho}\phi^{;\rho}\biggl)+ g_{\mu\nu}V(\phi)\biggl\}  + \frac{\omega}{\Phi^2}\biggr\{\Phi_{;\mu}\Phi_{;\nu} - \frac{1}{2}g_{\mu\nu}\Phi_{;\rho}\Phi^{;\rho}\biggl\} \nonumber \\
&+& \frac{1}{\Phi}\biggr\{\Phi_{;\mu;\nu} - g_{\mu\nu}\Box\Phi\biggl\},\\
\Box\Phi &=& - \frac{1}{3 + 2\omega}\biggr\{\epsilon \phi_{;\rho}\phi^{;\rho} - 4V(\phi)\biggl\},\\
\Box\phi &=& - \epsilon V_\phi(\phi).
\end{eqnarray}

The large scale structure of the universe is described by the FLRW flat metric,
\begin{eqnarray}
ds^2 = dt^2 - a(t)^2(dx^2 + dy^2 + dz^2).
\end{eqnarray}

The general equations of motion with a flat FLRW metric for a fluid  are,
\begin{eqnarray}
3H^2 &=& 8\pi\frac{\rho}{\Phi} + \frac{\omega}{2}\frac{\dot\Phi^2}{\Phi^2} - 3 H\frac{\dot\Phi}{\Phi},\\
\ddot\Phi + 3 H\dot\Phi &=& \frac{8\pi}{3 + 2\omega}(\rho - 3p),\\
\dot\rho + 3H(\rho + p) &=& 0.
\end{eqnarray}

In the case the self-interacting scalar field is used instead of a fluid, the equations are,
\begin{eqnarray}
\label{sfe1}
3H^2 &=& \frac{1}{\Phi}\biggr\{ \epsilon\frac{\dot\phi^2}{2} + V(\phi)\biggl\} + \frac{\omega}{2}\frac{\dot\Phi^2}{\Phi^2} - 3 H\frac{\dot\Phi}{\Phi},\\
\label{sfe2}
\ddot\Phi + 3 H\dot\Phi &=& - \frac{1}{3 + 2\omega}\biggr\{\epsilon\dot\phi^2 - 4 V(\phi)\biggl\},\\
\label{sfe3}
\ddot\phi + 3H\dot\phi  &=& - V_\phi(\phi).
\end{eqnarray}
Making an analogy with the fluid expressions, we can associate a energy density and a pressure to the scalar field:
\begin{eqnarray}
8\pi \rho_\phi &=& \epsilon\frac{\dot\phi^2}{2} + V(\phi),\\
8\pi p_\phi &=& \epsilon\frac{\dot\phi^2}{2} - V(\phi).
\end{eqnarray}
These expressions can be reverted, leading to
\begin{eqnarray}
\epsilon\dot\phi^2 &=& 8\pi (\rho_\phi + p_\phi),\\
V(\phi) &=& 8\pi (\rho_\phi - p_\phi),
\end{eqnarray}
As stated before, the violation of the null energy condition ($\rho_\phi + p_\phi < 0$) implies $\epsilon = - 1$.

\subsection{Jordan frame: solutions for a fluid}

In Ref. \cite{gurevich}, the authors have found the following general solutions for a perfect fluid with equation of state $p = \alpha\rho$ and $\omega > - 3/2$ in a flat FLRW metric:
\begin{eqnarray}
\label{gs1}
a(\eta) &=& a_0 (\eta - \eta_+)^{a_+}(\eta - \eta_-)^{a_-},\\
\label{gs2}
\Phi(\eta) &=& \Phi_0 (\eta - \eta_+)^{b_{+}}(\eta - \eta_-)^{b_-}.
\end{eqnarray}
In these expressions, $a_0$, $\Phi_0$, $\eta_\pm$ are constants. Moreover,
\begin{eqnarray}
a_\pm &=& \frac{\omega}{3\biggr(\beta \mp \sqrt{1 + \frac{2}{3}\omega}\biggl)},\\
b_\pm &=& \frac{1 \mp \sqrt{1 + \frac{2}{3}\omega}}{\beta \mp \sqrt{1 + \frac{2}{3}\omega}}
\end{eqnarray}
with,
\begin{eqnarray}
\beta = 1 + \omega(1 - \alpha).
\end{eqnarray}
The time $\eta$ is related to the cosmic time $t$ by,
\begin{eqnarray}
dt = a^{3\alpha}d\eta.
\end{eqnarray}

For $\omega < - 3/2$ the solutions read, 
\begin{eqnarray}
\label{gs1bis}
a &=& a_0 [(\eta + \eta_-)^2 + \eta_+^2]^\frac{\beta}{2A}\exp\biggr\{\pm\frac{\sqrt{\frac{2}{3}|\omega|}}{A}\arctan\frac{\eta + \eta_-}{\eta_+}\bigg\},\\
\label{gs2bis}
\Phi &=& \Phi_0 [(\eta + \eta_-)^2 + \eta_+^2]^\frac{(1 - 3\alpha)}{2A}\nonumber\\
&\times&\exp\biggr\{\mp 3(1 - \alpha)\frac{\sqrt{\frac{2}{3}|\omega| - 1}}{A}\arctan\frac{\eta + \eta_-}{\eta_+}\bigg\}, 
\end{eqnarray}
with $2A = ( 1- 3\alpha) + 3\beta(1 - \alpha)$, corresponding to a bouncing, singularity free, universe.

All these solutions (either for positive and negative $\omega$ reduces to the corresponding GR ones in the limit $|\omega| \rightarrow \infty$:
\begin{eqnarray}
a &\propto& t^\frac{2}{3(1 + \alpha)},\\
\Phi &\propto& \mbox{cte}.
\end{eqnarray}

If $\eta_\pm = 0$, the solutions (\ref{gs1},\ref{gs2}) can be written as,
\begin{eqnarray} 
a(\eta) &=& a_0 \eta^r,\\
\Phi(\eta) &=& \Phi_0 \eta^s,
\end{eqnarray}
with,
\begin{eqnarray}
r &=& a_+ + a_- = \frac{2[1 + \omega(1 - \alpha)]}{4 - 6\alpha + 3\omega(1 - \alpha)^2},\\
s &=& b_+ + b_- = \frac{2(1 - 3\alpha)}{4 - 6\alpha + 3\omega(1 - \alpha)^2}.
\end{eqnarray}

Using the relation between the time $\eta$ and the time $t$, we obtain,
\begin{eqnarray}
\eta = t^\frac{1}{1 + 3\alpha r},
\end{eqnarray}
leading to,
\begin{eqnarray}
a &=& \bar a_0t^\frac{r}{1 + 3\alpha r},\\
\Phi &=& \bar \Phi_0t^\frac{s}{1 + 3\alpha r}.
\end{eqnarray}
These expressions can be rewritten as,
\begin{eqnarray}
\label{sn1}
a &=& \bar a_0t^{\bar r},\\
\label{sn2}
\Phi &=& \bar \Phi_0t^{\bar s},
\end{eqnarray}
with,
\begin{eqnarray}
\label{sn3}
\bar r &=& \frac{2[1 + \omega(1 - \alpha)]}{4 + 3\omega(1 - \alpha^2)},\\
\label{sn4}
\bar s &=& \frac{2(1 - 3\alpha)}{4 + 3\omega(1 - \alpha^2)}.
\end{eqnarray}
Solution (\ref{sn1},\ref{sn2}), with definitions (\ref{sn3},\ref{sn4}) have been determined by Nariai \cite{nariai}. The solutions (\ref{sn3},\ref{sn4}) are valid for any value of $\omega$: both the solutions (\ref{gs1},\ref{gs2}) and (\ref{gs1bis},\ref{gs2bis}) imply (\ref{sn1},\ref{sn2}) in the limit $\eta_\pm = 0$.

Considering the fluid description of the matter/energy content and using the field equations, the following relation is obtained,
\begin{eqnarray}
\label{ratio}
8\pi \frac{\rho_0}{\Phi_0} = 3\biggr(\bar r + \frac{\bar s}{2}\biggl)^2 - \frac{(\omega + 3/2)}{2}\bar s^2.
\end{eqnarray}
The condition for gravity to be attractive is given by,
\begin{eqnarray}
\label{ine}
8\pi\frac{\rho_0}{\Phi_0} = 2\biggr(3 + 2\omega\biggl)\frac{[4 - 6\alpha + 3\omega (1 - \alpha)^2]}{[4 + 3\omega(1 - \alpha^2)]^2} > 0.
\end{eqnarray}
For $\alpha = 0$, the gravity is repulsive for $- 3/2 < \omega < - 4/3$, while for $\alpha = 1/3$ (radiation), it is positive for all $\omega$. \footnote{For $\alpha = 1/3$, the Nariai solutions reduce to the corresponding solution for a radiative universe in GR, with a constant gravitational coupling, and gravity is always attractive.} Two other interesting cases are $\alpha = 1$, for which the gravity is (surprisingly) attractive for $\omega < - 3/2$ and repulsive por $\omega > - 3/2$, and $\alpha = - 1$ (cosmological constant) for which gravity is repulsive for $- 3/2 < \omega < - 5/6$. 

Notice that $\rho$ and $\Phi_0$ are, for the moment, undetermined: only the ratio (\ref{ratio}) is known. We will consider $\Phi$ positive, since it is the requirement to perform the conformal transformation discussed below. 

\subsection{Jordan frame: scalar field representation}

If the sources are described by a self-interacting scalar field, the equations of motion are given by (\ref{sfe1})-(\ref{sfe3}).
We can rewrite equations (\ref{sfe1},\ref{sfe2}) as,
\begin{eqnarray}
\frac{\dot\phi^2}{\Phi} - 4\frac{V(\phi)}{\Phi} &=&  - (3 + 2\omega)\biggr\{\frac{\ddot\Phi}{\Phi} + 3H\frac{\dot\Phi}{\Phi}\biggl\},\\
 \frac{\dot\phi^2}{\Phi} + 2\frac{V(\phi)}{\Phi} &=& 6H^2 - \omega\frac{\dot\Phi^2}{\Phi^2} + 6H \frac{\dot\Phi}{\Phi}.
 \end{eqnarray}
 From these expressions, the scalar field and the potential are given by,
 \begin{eqnarray}
 \frac{\dot\phi^2}{\Phi} &=& 4H^2 - (2/3)\omega\frac{\dot\Phi^2}{\Phi^2} - \biggr(1 + \frac{2}{3}\omega\biggl)\frac{\ddot \Phi}{\Phi} + (1 - 2\omega)H\frac{\dot\Phi}{\Phi},\\
 \frac{V}{\Phi} &=& H^2 - \frac{\omega}{6}\frac{\dot \Phi^2}{\Phi^2} + \frac{1}{2}\biggr(1 + \frac{2}{3}\omega\biggl)\frac{\ddot\Phi}{\Phi} + \biggr(\frac{5}{2} + \omega\biggl)H\frac{\dot\Phi}{\Phi}.
 \end{eqnarray}

We use the expressions for $H$ and $\Phi$ from the previous section. Hence,
\begin{eqnarray}
\frac{\dot\Phi}{\Phi} &=& \frac{\bar s}{t},\\
\frac{\ddot\Phi}{\Phi} &=& \frac{\bar s(\bar s - 1)}{t^2},\\
H &=& \frac{\bar r}{t}.
\end{eqnarray}

In order these expressions represent the same solution of the field equations obtained previously using the hydrodynamical description, the scalar $\phi$ and the potential
$V(\phi)$ must behave as,

\begin{eqnarray}
\label{sj1}
\phi &=& \pm \phi_0 t^\frac{\bar s}{2}, \quad \phi_0 = \frac{\sqrt{32\pi(1 + \alpha)\epsilon\rho_0}}{\bar s},\\
\label{sj2}
V(t) &=& V_0 t^{(\bar s - 2)}, \quad V_0 = 4\pi(1 - \alpha)\rho_0.
\end{eqnarray}
We can write the potential in terms of the scalar field:
\begin{eqnarray}
\label{sj3}
V(\phi) = \bar V_0 \phi^{2(\bar s - 2)/\bar s}, \quad \bar V_0 = \frac{V_0}{\phi_0^{2/\bar s}}.
\end{eqnarray}
The potential follows a power law, while in GR it is an exponential \cite{Tomi}. 

The results above are valid for $\alpha \neq 1/3$. If $\alpha = 1/3$, $\bar s = 0$ and $\bar r = 1/2$. In this case, the scalar field and its potential are,
\begin{eqnarray}
\phi &=& \phi_0 \ln t,\\
V(\phi) &=& V_0 e^{-2\phi}.
\end{eqnarray}

\section{Conformal transformation and solutions in the Einstein frame}

The original Jordan frame discussed is characterized by a non-minimal coupling between gravity and the scalar field $\Phi$. It is possible to rewrite this Lagrangian in a minimal coupling framework (usually called Einstein frame) by performing the following conformal transformation \cite{wald}:
\begin{eqnarray}
g_{\mu\nu} = \Phi^{-1}\tilde g_{\mu\nu}.
\end{eqnarray}
Applying this transformation to the Lagangian, we obtain the Lagrangian:
\begin{eqnarray}
\label{lc}
{\cal L} = \sqrt{-\tilde g}\biggr\{\tilde R - \biggr(\omega + \frac{3}{2}\biggl)\frac{\Phi_{;\rho}\Phi^{;\rho}}{\Phi^2}\biggr\} + {\cal L}_m\biggr(\Phi\tilde g^{\mu\nu},\Psi\biggl), 
\end{eqnarray}
where $\Psi$ represents generically the matter fields.

Defining, 
\begin{eqnarray}
\sigma = \sqrt{|\tilde \omega|}\ln \Phi, \quad \tilde \omega = \omega + \frac{3}{2}, \quad \eta = \frac{\tilde\omega}{|\tilde\omega|},
\end{eqnarray}
the Lagrangian takes the form,
\begin{eqnarray}
\label{lcf}
{\cal L} = \sqrt{-\tilde g}\biggr\{\tilde R - \eta\sigma_{;\rho}\sigma^{;\rho} \biggl\} + {\cal L}_m\biggr(e^{- \frac{\sigma}{\sqrt{\tilde \omega}}}\tilde g_{\mu\nu},\Psi\biggl)
\end{eqnarray}

Next, this Lagrangian is studied in both the fluid and scalar field representations.

\subsection{Fluid representation}

The case where the matter fields are described by the fluid representation, the equations from the Lagrangian (\ref{lcf}) are given by,
\begin{eqnarray}
\label{fen1}
\tilde R_{\mu\nu} - \frac{1}{2}\tilde g_{\mu\nu}\tilde R &=& 8\pi G \tilde T_{\mu\nu} + \eta\biggr(\sigma_{;\mu}\sigma_{;\nu} - \frac{1}{2}\tilde g_{\mu\nu} \sigma_{;\rho}\sigma^{;\rho}\biggl),\\
\label{fen2}
\tilde{\Box}\sigma &=& 4\pi G\gamma\eta \tilde T,\\
\label{fen3}
{\tilde T^{\mu\nu}_{;\mu}} + \frac{\gamma}{2}\sigma^{;\nu}\tilde T &=& 0.
\end{eqnarray}
The notation $\gamma = 1/\sqrt{|\tilde\omega|}$ has been introduced in these expressions.
The energy-momentum tensor is given by,
\begin{eqnarray}
\tilde T_{\mu\nu} = \frac{T_{\mu\nu}}{\Phi^2} = e^{-2\gamma\sigma}T_{\mu\nu} =  (\tilde\rho + \tilde p)\tilde u_{\mu}\tilde u_\nu - \tilde p \tilde g_{\mu\nu}.
\end{eqnarray}
with,
\begin{eqnarray}
\tilde\rho = \rho e^{-2\gamma\sigma}, \quad \tilde p = p e^{-2\gamma\sigma},
\end{eqnarray}
with $\rho$ and $p$ the original energy density and pressure in the Jordan frame.

Using the flat FLRW metric the following equations of motion are obtained:
\begin{eqnarray}
\label{emn1}
3\tilde H^2 &=& 8\pi G\tilde\rho + \eta\frac{\dot\sigma^2}{2},\\
\label{emn2}
2\dot{\tilde H} + 3\tilde H^2 &=& - 8\pi G p - \eta \frac{\dot\sigma^2}{2},\\
\label{emn3}
\ddot\sigma + 3\tilde H(\tilde \rho + \tilde p) &=& 4\pi G \gamma\eta(\tilde \rho - 3\tilde p),\\
\label{emn4}
\dot{\tilde \rho} + 3\tilde H(\tilde\rho + \tilde p) + \frac{\gamma}{2}\dot\sigma(\tilde\rho - 3\tilde p) &=& 0.
\end{eqnarray}
In these expressions $\tilde H = \dot b/b$, where the dot now means derivative with respect to the cosmic time, denoted in the Einstein frame as $\tau$.
Supposing a linear equation of state $\tilde p = \alpha\tilde\rho$, power law solutions can be obtained:
\begin{eqnarray}
\label{sef1}
b &=& b_0 \tau^{\tilde r},\\
\label{sef2}
\sigma &=& \sigma_0 + \tilde s \ln \tau,
\end{eqnarray}
with,
\begin{eqnarray}
\label{es1}
\tilde r &=& \frac{4(1 - \alpha)\eta}{6(1 - \alpha^2)\eta + (1 - 3\alpha)^2\gamma^2},\\
\label{es2}
\tilde s &=& \frac{4(1 - 3\alpha)\gamma}{6(1 - \alpha^2)\eta + (1 - 3\alpha)^2\gamma^2}.
\end{eqnarray}
The energy density is given by,
\begin{eqnarray}
\tilde\rho &=& \tilde\rho_0 \tau^{- 3(1 + \alpha)\tilde r - \frac{(1 - 3\alpha)}{2}\gamma\tilde s}\nonumber\\
&=& \tilde\rho_0\tau^{-2}.
\end{eqnarray}
From (\ref{es1},\ref{es2}) and (\ref{emn1}), it can be obtained a relation for $\tilde\rho_0$:
\begin{eqnarray}
8\pi G\tilde\rho_0 = 3\tilde r^2 - \eta\frac{\tilde s^2}{2},
\end{eqnarray}
yielding,
\begin{eqnarray}
\label{ece}
	8\pi G\tilde\rho_0 = 2(3 + 2\omega)\frac{[4 - 6\alpha + 3\omega(1 - \alpha)^2]}{[5 - 3\alpha + 3\omega(1 - \alpha^2)]^2}.
\end{eqnarray}
Remark that the positivity of energy in the Einstein frame implies the same condition as in the Jordan frame, (\ref{ine}) since the only difference is in the denominator which is positive in both cases. This is expected since the conformal transformation preserves some general properties of the original frame like the causality. Moreover, the transformations made passing from one frame to another do not change the sign of energy density.

\subsection{Scalar field representation}

After a conformal transformation $g_{\mu\nu} = \phi^{-1}\tilde g_{\mu\nu}$ the Lagrangian takes the form given by (\ref{lc}). The field equations are given by (\ref{fen1}-\ref{fen3}), with the equations of motion for a FLRW universe given by (\ref{emn1}-\ref{emn4}).
In the scalar field representation the energy-momentum tensor is given by a self-interacting scalar field that mimics the fluid behavior at the background:
\begin{eqnarray}
8\pi G\tilde T_{\mu\nu} = \tilde\epsilon\biggr(\psi_{;\mu}\psi_{\nu} - \frac{1}{2}\tilde g_{\mu\nu}\psi_{\rho}\psi^{\rho}\biggl) + \tilde g_{\mu\nu}U(\psi),
\end{eqnarray}
with $\tilde\epsilon = \pm 1$ according the scalar field has positive (ordinary field) or negative (phantom field) kinetic energy.
Hence, for a background described by the FLRW metric, we find,
\begin{eqnarray}
8\pi G\tilde \rho &=& \tilde\epsilon\frac{\dot\psi^2}{2} + U(\psi),\\
8\pi G\tilde p &=& \tilde \epsilon\frac{\dot\psi^2}{2} - U(\psi).
\end{eqnarray}
These expression can be inverted, leading to,
\begin{eqnarray}
	\tilde\epsilon\dot\psi^2 &=& 8\pi G(1 + \alpha)\tilde\rho,\\
	U(\psi) &=& 4\pi G(1 - \alpha)\tilde\rho.
\end{eqnarray}
Notice that, since $\tilde\rho$ can be negative, $\tilde\epsilon = + 1$ if the combination $(1 + \alpha)\tilde\rho$ is positive, and $\tilde\epsilon = - 1$ if is negative.

By using the solutions in for the energy density found before, these relations leads to,
\begin{eqnarray}
\label{se1}
	\psi &=& \pm\sqrt{8\pi G\tilde\rho_0(1 + \alpha)\tilde\epsilon}\ln \tau,\\
	\label{se2}
	U &=& 4\pi G(1 - \alpha)\tilde\rho_0 \frac{1}{\tau^2}.
\end{eqnarray}
The potential $U$ can be expressed in terms of the scalar field $\psi$:
\begin{eqnarray}
\label{se3}
	U = 4\pi G(1 - \alpha)\tilde\rho_0 \exp\biggr\{\mp2\frac{\psi}{\sqrt{8\pi G \tilde\rho_0(1 + \alpha)\tilde\epsilon}}\biggl\}.
		\end{eqnarray}
		The constant $\rho_0$ is given, in terms of $\alpha$ and $\omega$, by (\ref{ece}) 
		
		The results above can be compared with those of Ref. \cite{Tomi} by taking the limit $\omega \rightarrow \infty$ which is equivalent to $\gamma \rightarrow 0$. In this limit, $\tilde r \rightarrow 2/3(1 + \alpha)$ and $\tilde s \rightarrow 0$, leading to,
		\begin{eqnarray}
		8\pi G\tilde\rho_0 = \frac{4}{3(1 + \alpha)^2}, 
		\end{eqnarray}
		implying,
		\begin{eqnarray}
	\psi &=& \pm\frac{2}{\sqrt{3(1 + \alpha)\tilde\epsilon}}\ln \tau,\\
	\label{pef}
	U &=& \frac{2}{3}\frac{1 - \alpha}{(1 + \alpha)^2} \exp\biggr\{\mp \sqrt{3(1 + \alpha)\tilde\epsilon}\psi\biggl\}.
\end{eqnarray}
These are the solutions found in Ref. \cite{Tomi}, with the addition of the parameter $\tilde\epsilon$ in order to take into account phantom fluids represented by $\alpha < - 1$. The connection with the expressions found for the scalar representation 
in the Jordan frame discussed in section 3 can be made easily. Remark however that transposing (\ref{se1}-\ref{se3}) to the Jordan frame does not lead to the same expressions as (\ref{sj1}-\ref{sj3}) due to the redefinitions made in the matter fields which results in equations (\ref{fen1}-\ref{fen3}).

\section{Analysis of the results}

We now investigate the conditions for an accelerated expansion scenario in the late and early universe. The transitions between decelerated/accelerated (and vice-versa) expansions will be identified. The scalar field potential for each possible scenario will be discussed. The goal is to identify configurations which could describe the effects that could be  either attributed to dark matter and dark energy, mainly due to the dynamical transitions from a dark matter dominated universe and dark energy dominated universe. This transition is driven dynamically by the gravitational coupling variability since the matter content contains, for each case, a single fluid with a fixed equation of state with a constant relation between pressure and energy density. 

The many possible scenarios given by the Gurevich solutions are, in principle, determined by the set of parameters $\alpha$ and $\omega$, which are quite convoluted. Hence, we will exemplify some of the different possibilities by considering tree values for the equation of state parameter: $\alpha = 0$, $-1$ and $-2$. This last case represents a deep phantom matter. As it will emerge from the analysis below, the radiative case $\alpha  = 1/3$ predicts an always decelerating universe. Those different cases will be analyzed in the Jordan and Einstein frames.

\subsection{Jordan frame}

In the Jordan frame the solution for the scale factor $a$ is given by (\ref{sn3}), while the gravitational coupling is related to the inverse of $\Phi$ which is given by (\ref{sn4}). There are two possible accelerated regimes describing an expanding universe:
$\bar r > 1$ ($ 0 \leq t < \infty$)  or $\bar r < 0$ ($ -\infty < t \leq 0$). First, we consider the late time universe for which the constants $\eta_\pm$ can be neglected and the Gurevich's solution reduces to the Nariai one.

\begin{itemize}

\item $\alpha = 0$ (pressureless matter).

Fixing a pressureless matter, the late time solutions are,
\begin{eqnarray}
a &\propto& t^\frac{2 + 2\omega}{4 + 3\omega},\\
\Phi &\propto& t^\frac{2}{4 + 3\omega},
\end{eqnarray}
and the two possible accelerated regime are given by:
\begin{enumerate}
\item $\bar r > 1$ ($ 0 \leq t < \infty$). The accelerated expansion is obtained when $- 2 < \omega < - 4/3$. In this case, the energy density is negative in the interval $- 3/2 < \omega < - 4/3$ and positive in the interval $- 2 < \omega < - 3/2$, see relation (\ref{ine}). The gravitational coupling is increasing;
\item $\bar r < 0 $ ($ -\infty < t \leq 0$). The inflationary regime occurs for $- \frac{4}{3} < \omega < - 1$, corresponding to a positive energy density and the gravitational coupling is increasing. 
\end{enumerate}

The potential is given, in terms of $\phi$, as,
\begin{eqnarray}
V(\phi) = V_0 \phi^{-6(1 + \omega)}.
\end{eqnarray}
It increases monotonically with $\phi$ if $\omega < - 1$, and it decreases monotonically if $\omega > - 1$.

\item $\alpha = - 1$ (cosmological constant). 

For a cosmological constant the solutions, for a late time universe, are
\begin{eqnarray}
a &\propto& t^{\omega + \frac{1}{2}},\\
\Phi &\propto& t^2.
\end{eqnarray}
These solutions have been discussed, in the context of the extended inflationary scenario, in Ref. \cite{stein}.
The two possible accelerate expansion are given by,
\begin{enumerate}
\item $\bar r > 1 \quad \rightarrow  \quad \omega > 1/2$. The energy density is always positive;
\item $\bar r < 0 \quad \rightarrow \quad \omega < - 1/2$. The energy is positive for $ - \infty < \omega < - 3/2$ and $ - 5/6 < \omega < - 1/2$. Energy is negative if $  -3/2 < \omega < - 5/6$.
\end{enumerate}
As expected, the potential is constant.

\item $\alpha = - 2$ (phantom matter).

The equation of state corresponding to the phantom regime is also relevant, since the analysis of the cosmological observational data indicates the possibility dark energy has a phantom character. For instance, if $\alpha = - 2$, the solutions for the scale factor and the gravitational coupling are,

\begin{eqnarray}
a &\propto& t^\frac{2 + 6\omega}{4 - 9\omega},\\
\Phi &\propto& t^\frac{14}{4 - 9\omega}.
\end{eqnarray}

 Accelerated expansion is possible if,
 
\begin{enumerate}
 \item $  \bar r > 1 \quad \rightarrow \quad 2/15 < \omega<4/9$. Energy density is always positive;
 \item  $ \bar r < 0 \quad \rightarrow \quad \omega < - 1/3$ or $\omega > 4/9$. Energy is always positive except for $- 3/2 < \omega < - 16/27$.
\end{enumerate}

The potential is given, in terms of $\phi$, as,
\begin{eqnarray}
V(\phi) = V_0 \phi^{6\frac{(1 + 3\omega)}{7}},
\end{eqnarray}
and it increases with $\phi$ for $\omega > - 1/3$ and decreases if $\omega < - 1/3$.
\end{itemize}

\subsection{Einstein frame}

\begin{itemize}
\item $\alpha = 0$.

In the Einstein frame the solution for the scale factor $b$, with $\alpha = 0$, is given by
\begin{eqnarray}
b \propto \tau^\frac{4\omega}{1 + 6\omega}.
\end{eqnarray}
The gravitational coupling is constant. Again, there are two possible accelerated regimes for an expanding universe according
$\tilde r > 1$ ($ 0 \leq \tau < \infty$)  or $\tilde r < 0$ ($ -\infty < \tau \leq 0$).

\begin{enumerate}
\item $\tilde r  > 1$ ($ 0 \leq \tau < \infty$). The accelerated regime is obtained when $-  1/2  < \omega < - 1/6$. In this case, the energy density is positive, see relation (\ref{ece}).
\item $\tilde r < 0 $ ($ -\infty < \tau \leq 0$). The accelerated regime occurs for $- 1/6 < \omega < 0$, corresponding again to positive energy density. 
\end{enumerate}

The potential is given by,
\begin{eqnarray}
U(\psi) = \frac{(3 + 2\omega)(4 + 3\omega)}{(5 + 3\omega)^2}\exp\biggr\{ \mp \frac{5 + 3\omega}{\sqrt{2(3 + 2\omega)(4 + 3\omega)}}\psi\biggl\}.
\end{eqnarray}
The potential has two branches, exponentially increasing or decreasing with $\psi$.


\item $\alpha = - 1$.

For the cosmological constant case, the solution for the scale factor in the Einstein frame reads,
\begin{eqnarray}
b \propto \tau^\frac{\omega}{2}.
\end{eqnarray}
Again, accelerated expansion are possible in two cases:
\begin{enumerate}
\item $\omega/2 > 1$ ($0 < \tau < \infty$), implying $\omega > 2$. with positive energy;
\item $\omega/2 < 0$ ($- \infty < \tau < 0$), implying $\omega < 0$, with negative energy except in the interval $ - 5/6 < \omega < 0$.
\end{enumerate}

The potential is constant.

\item $\alpha = - 2$.

For this case, the scale factor behave as,
\begin{eqnarray}
b \propto \tau^\frac{12\omega}{49 - 18\omega}.
\end{eqnarray}
The accelerated expansion occurs if,
\begin{enumerate}
\item  $\tilde r > 1\quad \rightarrow\quad 49/30 < \omega < 49/18$,\\
\item  $\tilde r <  0 \quad \rightarrow\quad \omega < 0 \text{ or }  \omega > 49/18$.
\end{enumerate}
The energy is positive when $\omega > - 27/16$ or $\omega < - 3/2$.

The potential reads,
\begin{eqnarray}
U(\psi) = \frac{3(3 + 2\omega)(16 + 27\omega)}{(11 - 9\omega)^2}\exp\biggr\{ \mp \frac{11 - 9\omega}{\sqrt{6(3 + 2\omega)(16 + 27\omega)}}\psi\biggl\}.
\end{eqnarray}
Again, it has two branches according the sign in the exponential.

\end{itemize}

\subsection{Behavior in the primordial universe}

The general solutions displayed above for the Brans-Dicke theory has a curious behavior as the initial singularity is approached, $\eta \rightarrow \eta_+$ in (\ref{gs1},\ref{gs2}): the matter component becomes irrelevant and the universe approaches a BD vacuum state.
In the Jordan frame, the scale factor evolves, near the singularity for $\omega > - 3/2$ as,
\begin{eqnarray}
\label{vs}
a \sim t^\frac{1 + \omega + \sqrt{1 + \frac{2}{3}}}{4 + 3\omega},
\end{eqnarray}
which represents a decelerating universe. For $\omega < - 3/2$ the solution corresponds to a bouncing universe, exhibiting an accelerated phase $\ddot a > 0$ near the bounce. 

In the Einstein frame the asymptotic initial solution is typical of a stiff matter source,
\begin{eqnarray}
b \sim \tau^\frac{1}{3},
\end{eqnarray}
what is expect since the matter component become subdominant and the only source is a free scalar field $\sigma$.

Hence, the solution discussed in this work shows many possibilities of an initial decelerating universe evolving to an accelerating one. In many cases the matter energy density is positive. This transition decelerated/accelerated is driven by the dynamical gravitational coupling. Remark that the being vacuum solutions, there is no self-interacting scalar field associated to the cosmic evolution in primordial time: such self-interacting scalar field emerges as the universe evolves deviating from the vacuum configuration.

\section{Conclusions}

A given solution obtained in a gravity theory using a hydrodynamical can be reproduced using instead a self interacting scalar field with an appropriate potential. From the background point of view, there is no essential difference in using one or another approach. However, huge differences may emerge at background level. One of the reason is the sound velocity associated with one or other representation: for the hydrodynamical approach, the velocity depends on the equation of state and of the adiabatic or non-adiabatic nature of the perturbations; for a canonical self interacting scalar field, the velocity is always equal to one (in units of $c$), see Ref. \cite{neven}. The discussion of the hydrodynamical and scalar field representation, at the background and perturbative levels, in General Relativity has been carried out, in the cosmological context, in Ref. \cite{Tomi}. 

We have extended this discussion,  here, to the Brans-Dicke scalar-tensor theory, perhaps the simplest theory with a non-minimal coupling between gravity and the scalar field. However, we have restricted ourselves to the background analysis. The reason is that, frequently, in studying the BD theory, it is more convenient to perform a conformal transformation breaking the non-minimal coupling, and recovering a frame where gravity is coupled minimally to a scalar field. The price to pay is that the energy-momentum tensor interacts non-trivially with the scalar field. 

The issue of mapping a solution in one frame to a solution in the other frame makes the analysis more complex. Moreover, at perturbative level, the definition of the perturbed quantities is affected by the conformal transformation in a non-trivial way. For these reasons, the perturbative analysis is postponed to future work. However, some considerations on the sound velocity may anticipate that unstable solutions in the hydrodynamical representation may be mapped into stable solutions in the scalar-tensor representation.

While in GR the potential reproducing a given hydrodynamical solution follows an exponential function of the scalar field, in BD theory it is a power law. In both cases the potential has a monotonic behavior as function of the scalar field or, alternatively, of time. In the Einstein frame formulation of the BD theory, the exponential potential is recovered. The specific expressions for the behavior of the scale factor, scalar field and the potential are functions of the Brans-Dicke parameter $\omega$ and the equation of state parameter $\alpha$. A more detailed analysis has been made for three specific cases: zero pressure fluid, the cosmological constant and a particular phantom component. It has been shown that in both Jordan and Einstein frames accelerated expansion is obtained either with positive or negative energy but for different ranges of the BD parameter $\omega$. Hence, an accelerated (decelerated) expansion may be mapped into a decelerated (accelerated) expansion in another frame.

One interesting aspect of the classes of solutions discussed here is that the primordial evolution corresponds to the vacuum state, in both JF and EF, exhibiting only decelerated expansion. Hence, the fluid or self-interacting scalar field becomes efficient only for the late universe, determining its ulterior evolution, and in may cases implying a transition to an accelerated regime. 

Two last remarks. Firstly, the constraints coming from the PPN analysis imply that the BD parameter $\omega$ may be quite huge \cite{will} even if the precise estimations depend if the analysis is made in the JF or in EF \cite{ppn}. In both cases, the PPN tests indicate that $\omega$ must of the order or exceed 40.000. However, some of the most interesting solutions displayed here, mainly for $\alpha = - 1$ and $\alpha = - 2$, show the transition deceleration/acceleration for very large $|\omega|$. A more complex configuration, with a $\omega$ function of $\Phi$ instead of a constant, may also reconcile some scenarios here (as those with $\alpha = 0$) with the PPN estimations. See for example Ref. \cite{jdb} for a description of such class of extension of the BD theory. Secondly, the importance of the field representation, besides become clear the connection of the components of the dark sector with fundamental frameworks, like string and grand unified theories, resides in the formation of structure problem. As an example, in the the case $p = - \rho$, while the background implies an homogeneous scalar field in the matter/energy sector, the local fluctuations excites the scalar model, that will contribute to the evolution of the perturbations, leading also to the phenomena of dark energy clustering in opposition to what happens in the strict fluid representation, see Ref. \cite{pbd}. The formation of cosmic structures will be addressed in a separate work.

\vspace{0.5cm}

\noindent
{\bf Acknowledgements:} We thank CNPq (Brazil) and FAPES (Brazil) for partial financial support. T.H.B.A. thanks also CAPES for a fellowship during the elaboration of this work. L.F.G. also thanks Fundação Araucária for financial support.

\end{document}